\documentclass[11pt]{article}
\usepackage{moriond,epsfig}

\def\PRL{\em Phys. Rev. Lett.}
\def\PRD{{\em Phys. Rev.} D}

\def\be{\begin{equation}}
\def\ee{\end{equation}}
\def\bea{\begin{eqnarray}}
\def\eea{\end{eqnarray}}

\begin{document}
\vspace*{4cm}
\title{HIGH ENERGY EVOLUTION WITH POMERON LOOPS}

\author{ M. LUBLINSKY }

\address{Department of Physics, University of Connecticut,\\
2152 Hillside Rd, Storrs, CT 06269, USA}

\maketitle\abstracts{The high energy/density QCD has been widely used for DIS phenomenology
with a projectile particle  considered as perturbative and dilute.
 We review some recent attempts to derive a high energy evolution kernel which
 treats targets and projectiles in a symmetric manner. From 
 theoretical  point of view the problem is tightly related to inclusion of Pomeron
 loops in the evolution. The ultimate goal is to consider high
  energy scatterings with both projectile and target being dense, the
 situation faced at RHIC and the LHC. 
}

\begin{figure}
\centerline{\psfig{figure=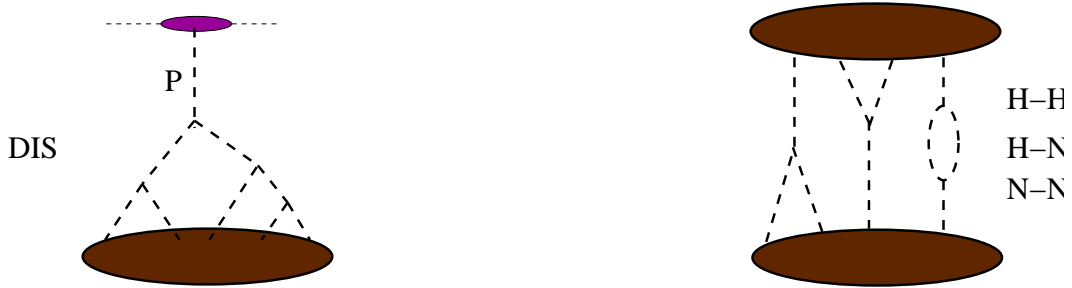,height=1.5in}}
\caption{DIS vs Hadron-Hadron collision.
\label{fig1}}
\end{figure}

Thanks to the  diluteness of perturbative projectile,
the high energy limit of DIS is understood relatively well. Recently the main theoretical effort has been shifted 
towards high energy scattering of dense objects, which could be hadrons and/or nuclei. The theory has phenomenological
implications for the LHC, RHIC and TeVatron.  The physics of DIS is linear in projectile`s density but non-linear
in the density of a dense target. At high energies we resum Pomeron fan diagrams (Fig. \ref{fig1}) using GLR-type
non-linear evolution equations. The modern version of the GLR equation is known as the JIMWLK evolution. In contrast to DIS,
in hadron-hadron collisions physics is non-linear both in projectile`s and target`s densities. We have to resum both
up and down type Pomeron fan diagrams as well as Pomeron loops (Fig. \ref{fig1}). 
We have been able to achieve some modest progress in understanding high energy evolution with Pomeron loops 
\cite{MS,dipole,kl,HIMST}.
New ideas include an extension of the JIMWLK equation, known as JIMWLK+; an evolution equation in the dilute limit, the -
KLWMIJ kernel; a new symmetry transformation - the Dense Dilute Duality (DDD) as well as the prove of Self-Duality 
of the complete and yet unknown evolution kernel \cite{kl}. Several new results have been obtained in the dipole evolutions 
with Pomeron loops \cite{dipole},  but these developments will not be reviewed in this talk.

\begin{figure}
\centerline{\psfig{figure=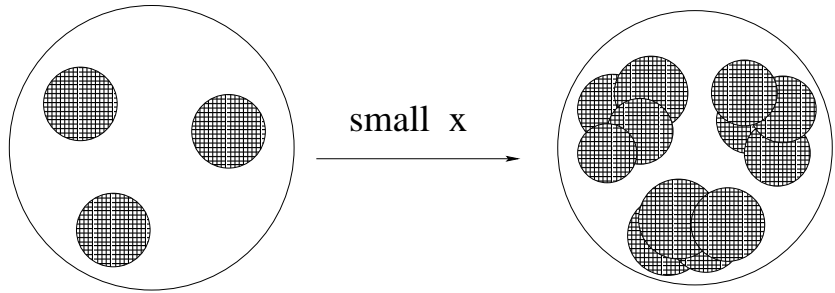,height=1.5in}}
\caption{BFKL evolution.
\label{fig2}}
\end{figure}

Let us start from considering high energy evolution of  hadron`s wavefunction. In the dilute limit, the wavefunction
consists of a few valence partons (gluons). When boosted by a small amount, the valence partons emit gluons (Fig. \ref{fig2})
and the probability of emission is proportional to the number of emitters: $\delta\rho\sim \rho$.  At finite rapidity,
consequent emissions are summed up and lead to an exponential growth of the density with energy: $\rho\sim e^{c Y}$, 
the BFKL result. 
The density cannot grow exponentially as this would eventually violate unitarity. At high densities
a new mechanism taming the growth widely known as saturation effects  is needed.

\begin{figure}
\psfig{figure=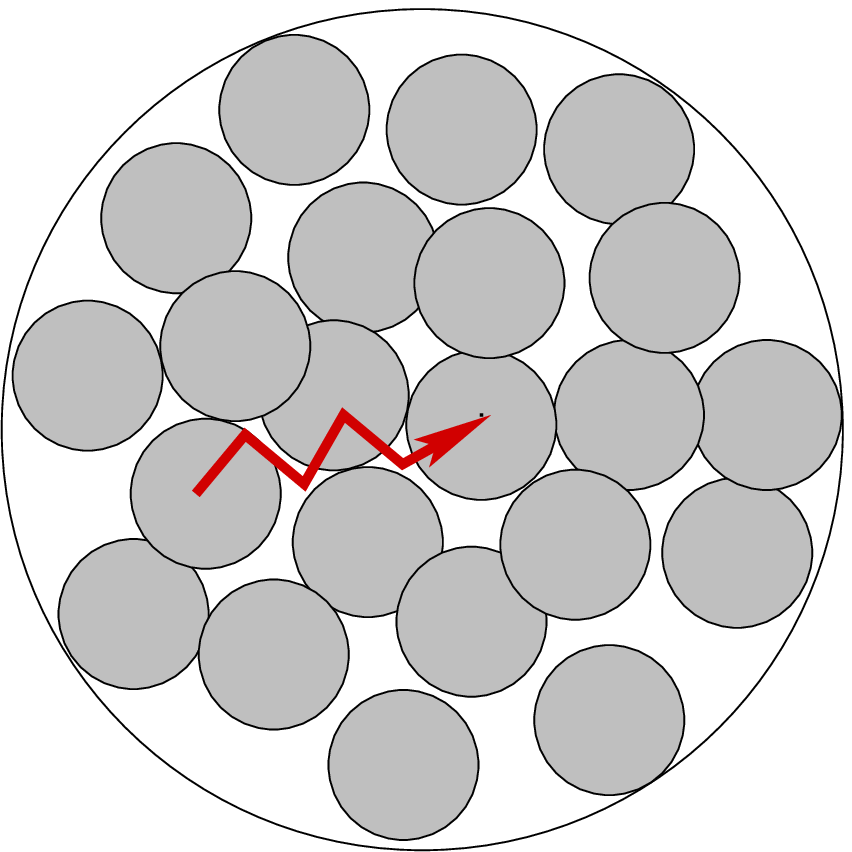,height=1.5in}
\hspace{1.5cm} \psfig{figure=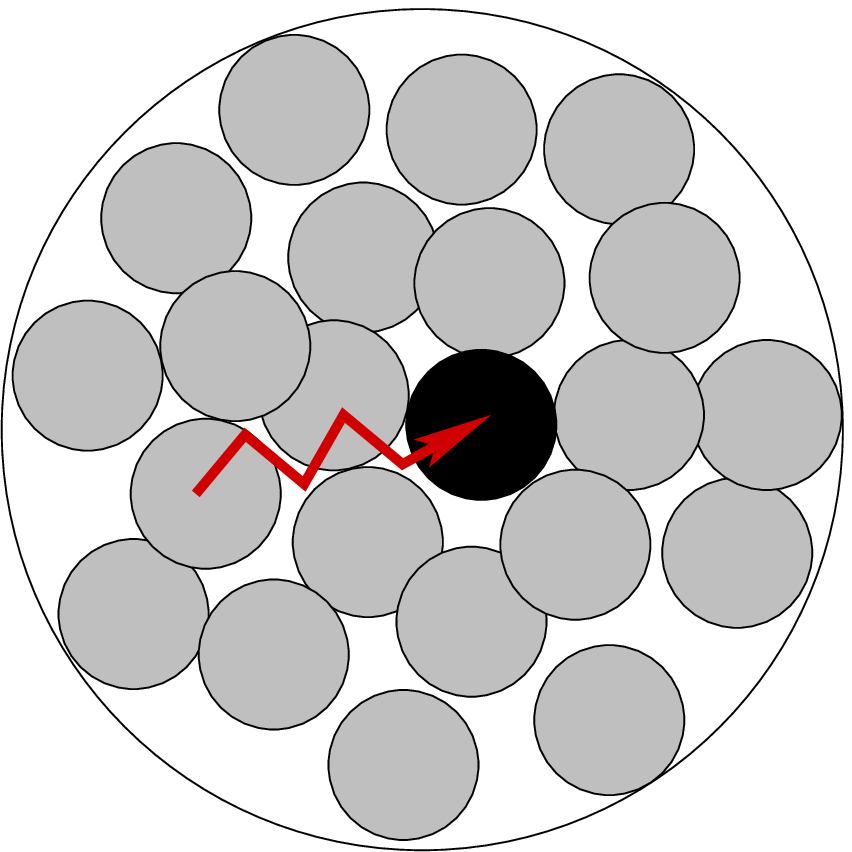,height=1.5in}
\hspace{1.5cm}\psfig{figure=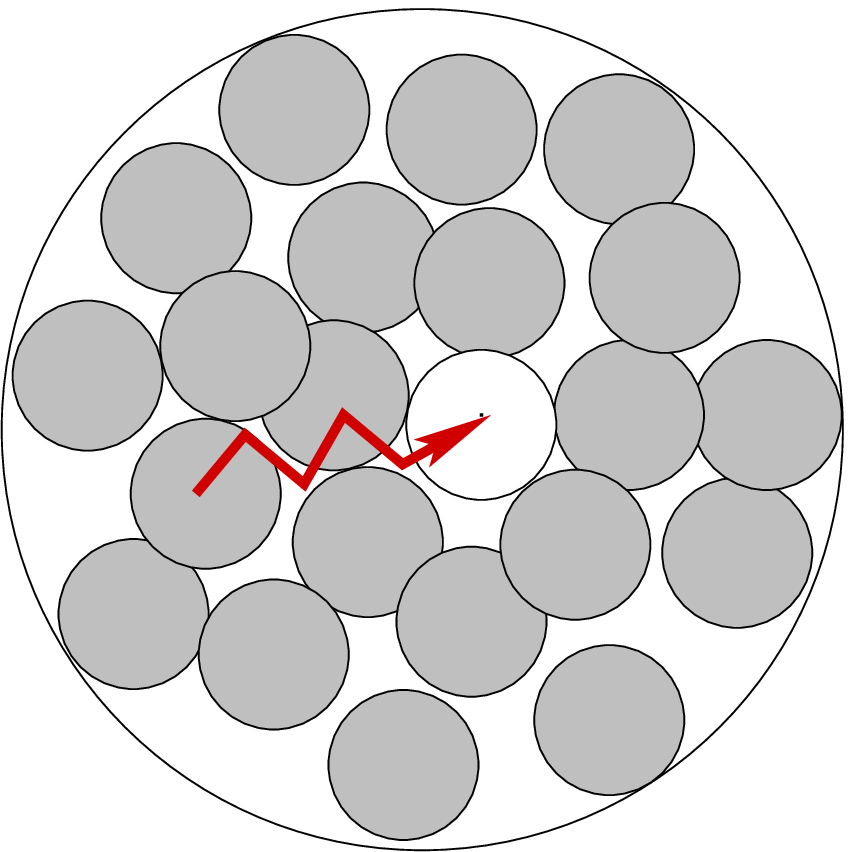,height=1.5in}
\caption{Random walk and color bleaching. 
\label{fig3}}
\end{figure}

What actually happens in the dense regime. We expect two phenomena. First, the emission probability would be independent
of the density. Second, a phenomenon of ``color bleaching'' should become important (Fig. \ref{fig3}).
 As a result, we expect  
a random walk with the density $\rho$ growing as $\sqrt Y$. 
The color bleaching appears 
when a gluon is emitted  in a pont where another gluon is already present. Adding color representations, this point
can either turn into blacker or more absorptive for an external probe, or become  a ``hole'' with zero net color 
charge. 

The dilute phase with a few gluons and the dense phase with a few holes are in fact related by the
Dense Dilute Duality transformation. The underlying Reggeon Field Theory reveals a structure with two degenerate
vacua: the absolutely empty ``white'' vacuum and fully absorptive ``black'' one. Gluons can be viewed as excitations
over the white vacuum, while holes are excitations over the black one. The spectrum of these excitations is degenerate.

Let us now focus on high energy scattering. A general expression for $S$-matrix is given by the quantum-mechanical
expectation value of $S$-matrix operator sandwiched over direct product of target and projectile wavefunctions:
$$S(Y)\,=\,\langle P \,\,\langle T|\,\hat S(\rho^t,\,\rho^p)\,|T\rangle_{Y_0} \,P\rangle_{Y-Y_0}\,. $$
The target is assumed to be evolved to rapidity $Y_0$ while the projectile takes on itself the rest of the total rapidity
$Y$. The $S$-matrix has to be independent of $Y_0$.  

\begin{figure}
\centerline{\psfig{figure=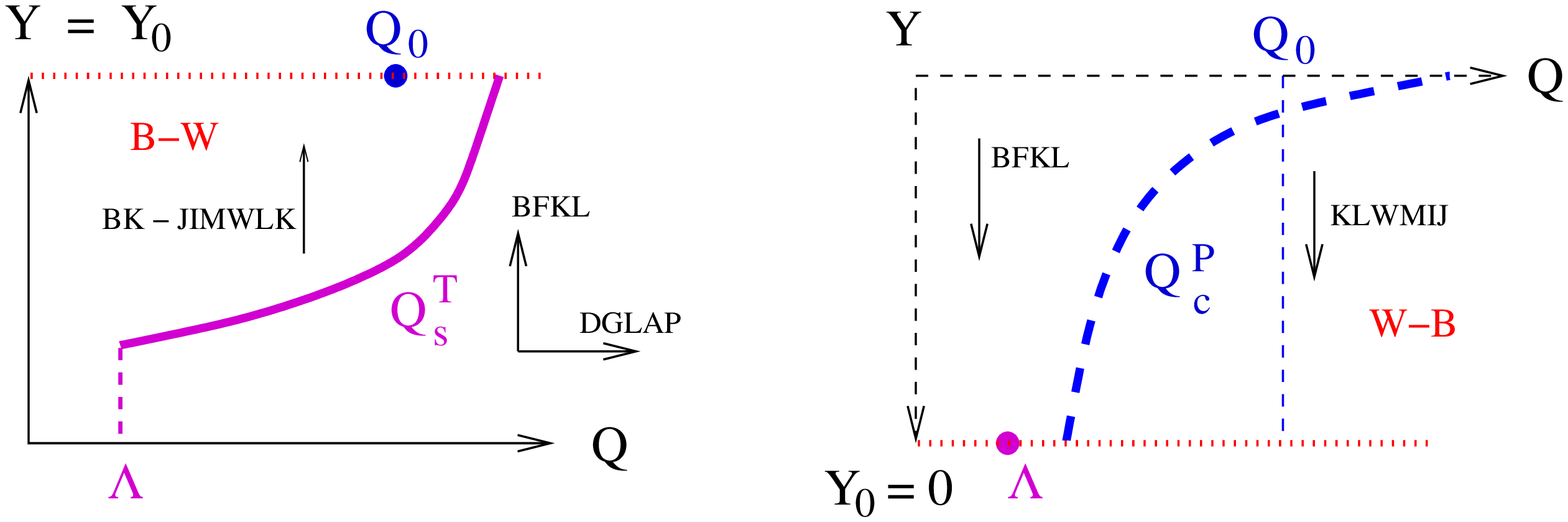,height=2in}}
\caption{DIS: a) - projectile`s rest frame. b) - target rest frame.
\label{fig4}}
\end{figure}

Consider DIS. Fig. \ref{fig4},a presents a standard picture of DIS in the projectile`s rest frame. The dense target evolves
with energy  developing a  saturation scale $Q_s^T$. This scale separates linear evolution from
non-linear  and can be associated with a solution of the JIMWLK equation. At the end, this evolved wavefunction
is probed by a virtual photon $Q_0$.  If $Q_0$ is within the vicinity of $Q_s^T(Y)$ then the cross section depends
on these two scales only. As a result, $F_2=F_2(Q_0/Q_s^T)$ known as the geometrical scaling. 
Here is the standard estimate for the saturation scale 
$$\rho^t\,\simeq\,\Lambda^2\,e^{c\,Y}\,;\quad\quad\quad\quad\quad\quad \alpha_s\,\rho^t/Q_s^2\,\sim\,1$$  
Suppose now that instead of evolving the target we would evolve the dilute projectile keeping the target at rest.
How would the projectile evolve? Since it is dilute, we would naively say it evolves a la BFKL.  This is incorrect
as in this case we would obtain a result different from the previous case. A correct projectile`s evolution has to reflect
the non-linear dynamics discussed above and should also generate a scale. Indeed, a dilute probe evolves according  
to a non-linear evolution, which we refer to as KLWMIJ \cite{kl} (Fig. \ref{fig4},b). A new scale $Q_c^P(Y)$ is 
associated with its solution. Th physical meaning of $Q_c^2$ is that of the density of a potential target which when
scattered on the given projectile would be absorbed with probability of order one \cite{MS}: 
$$\sigma\,=\,\alpha_s\,\rho^p\,Q_c^2/Q_0^4\,\sim\,1\,;\quad\quad\quad\quad\quad\quad \rho^p\,\simeq\,Q_0^2\,e^{c\,Y}\,.$$

The KLWMIJ kernel (Hamiltonian) is in fact a twin brother of the JIMWLK and is related to it by the
 Dense Dilute Duality (DDD) transformation \cite{kl}. Within the eikonal approximation the DDD transformation is just a
functional Furier transform:  
$$\chi^{KLWMIJ}[\rho^p]\,=\,\int_{\alpha^t} \chi^{JIMWLK}[\alpha^t]\,e^{i\,\rho^p\,\alpha^t}
$$
with $\alpha^t$ being a field created by the sources in the target: 
$\Delta\,\alpha^t\,\,''=''\,\,\rho^t \,\,\,\,\,\,(YME)$.

\begin{figure}
\centerline{\psfig{figure=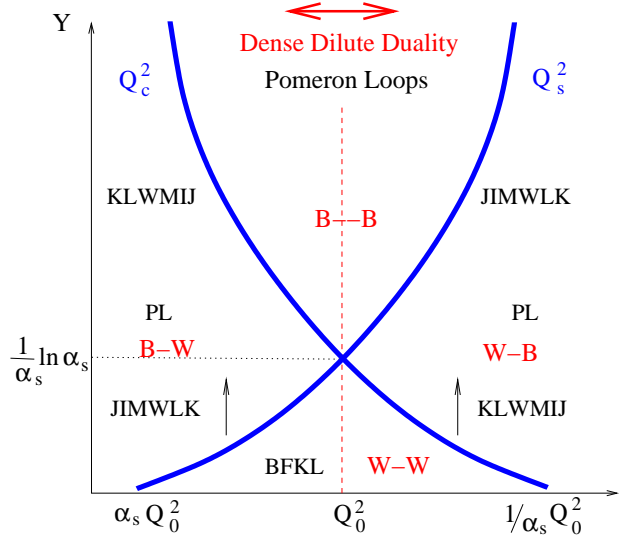,height=2.8in}}
\caption{High energy evolution of a dilute probe.
\label{fig5}}
\end{figure}

The following message is most important: any wavefunction, depending how it is probed, would display both
dense and dilute properties. Consequently, figures \ref{fig4},a and \ref{fig4},b have to be combined to describe evolution
of a single wavefunction. Fig. \ref{fig5} displays evolution of the projectile. It has the same KLWMIJ evolution as before.
However, as we further evolve with rapidity, even initially dilute wavefunction becomes dense and develops its
own saturation scale. The relevant evolution then becomes that of the JIMWLK.  The appearance of two scales in the evolution
is expected to result in violation of the geometrical scaling \cite{MS}.
 In one of the regions the linear
BFKL evolution is still valid.  
In the rest of the regions Pomeron loops are important. 
The JIMWLK and KLWMIJ evolutions represent the dense and dilute limits of a complete kernel.  
Any interpolation between these two limits would necessary generate Pomeron loops.

A symbolic mirror symmetry across the vertical line reflects the DDD symmetry  mentioned above. In fact we were
able to prove that under the DDD transformation the complete evolution kernel $\chi$ must be self-dual \cite{kl}:
$$
\chi(i\,\alpha\,\,,\delta/\delta\,\alpha)\,\,\,=\,\,\,\chi(\delta/\delta\,\rho,
\,\,i\,\rho)$$
It is likely that this self-duality condition is equivalent to the $t$-channel unitarity of the theory.

\section*{Acknowledgments}
All of the results presented were obtained in collaboration with Alex Kovner.
I am most grateful to Alex  for being a wonderful collaborator. I would like to thank Larry McLerran 
for the invitation to Moriond. The support of the organizing committee is  thankfully acknowledged.

\end{document}